\documentclass{article}
\usepackage[dvips]{graphicx}
\usepackage{amsmath,amssymb}

\begin{document}
\title{ Macroscopic Interferences of Neutrino Waves} 
\author{ K. Ishikawa and  Y. Tobita\\
\\
  Department of Physics, Faculty of Science, \\
Hokkaido University Sapporo 060-0810, Japan
}
\maketitle
\begin{abstract}
 Interference phenomena of neutrinos are studied. High energy
neutrino in T2K near detector and low energy neutrino in KamLAND are
possible experiments that could show macroscopic interferences of
neutrino waves. In both experiments interference patterns may give new
insights on the absolute value of the neutrino mass. 
\end{abstract}
\begin{center}
\end{center}


\section{Interference of neutrino waves}
If a wave function $\psi_{\nu}(t,{\vec x})$ is  a sum of
two different wave functions, 
the probability of observing this   particle 
that is proportional to $|\psi(t,{\vec x})|^2$ 
has an  interference term. Interference phenomena of the light,
electron, and neutron have been
studied well and are used for many purposes in wide area. The neutrino is a 
wave and interacts with matters so weakly that it is extremely difficult to
control the neutrino. It is a challenging task  to prepare   a detection
method  for the neutrino interference.

\begin{figure}[t]
 \centering
 \includegraphics[angle=00, scale=.6]{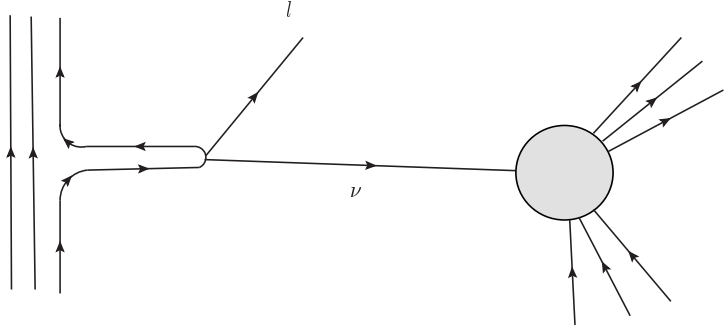}
 \caption{The whole processes of the neutrino reaction is shown. The
 neutrino is produced as a decay product of the pion or nucleon and
 propagates for macroscopic distance. The neutrino is detected finally. }
\label{neutrino-reaction-diagram}
\end{figure}

In this article, we show that 
diffraction-like experiments of neutrinos are 
possible despite its extreme weakness of interactions  with matters 
\cite{Ishikawa-Tobita-ptp}\cite{Ishikawa-Tobita-ptp2} and that they give new insights on neutrinos. Due
to the weakness of the interactions it is necessary  to analyze whole 
processes from its production to detection \cite{Ishikawa-Shimomura}.  
Neutrino becomes  a wave packet of a finite
coherence length and  behaves as a wave and
particle simultaneously. Its coherence properties are important for the
observability of the interferences. 
Neutrino is produced in the scattering or decay of a particle
by  the weak interaction  and is in the intermediate state
between  the production and 
detection processes\cite{Asahara} as is shown in Fig. \ref{neutrino-reaction-diagram}, 
\begin{eqnarray}
particle+target \rightarrow neutrino \rightarrow neutrino +target 
 \rightarrow particles.
\end{eqnarray}
For the interference pattern of the neutrino to become observable,  a wave at
the detector should be  a superposition of multiple waves of different phases.  

Distance, $L$, and energy, $E_{\nu}$, of the neutrino propagations of  T2K \cite{K2K} 
near detector 
and KamLAND \cite{KamLAND-Reactor}, which we focus in the present work are
\begin{eqnarray}
& &\text{T2K}\,:\,L=few \times 100\,[\text{m}].;~  E_{\nu}= few \times [\text{GeV}/c], \\
& &\text{KamLAND}\,:\,L= 200 [\text{km}]. ;~ E_{\nu}=few \times [\text{MeV}/c]. 
\end{eqnarray}
$L$ is the distance between the reactor and detector in KamLAND, and is
shorter than the distance between pion source and detector in T2K, since
 pions decay in large decay area. These parameters are different each
 others but two experiments show
similar phenomena of interferences.

\section{Neutrino production and detection :wave packet scattering }

The amplitude of the neutrino reactions from its production to detection 
is written as 
\begin{eqnarray}
\langle l,\alpha_{out},\beta_{out}|\int dx_1 dx_2
 T(H_i(x_1)H_i(x_2))|\alpha_{in},\beta_{in} \rangle,
\end{eqnarray} 
with a weak interaction Hamiltonian $H_i(x)$ and the in-states
 $\alpha_{in},\beta_{in}$ and out-states $\alpha_{out},\beta_{out}$ and
 a lepton $l$.
In the standard S-matrix, the in-states and out-states are plane waves 
defined at
 $t=-\infty$ and at $t=\infty$. 
Hence this amplitude is invariant under the translation of the space and
 time and   is useless for the space-time dependent informations in 
scatterings such as  probability at a finite
time interval or finite distance.    
These translational
non-invariant quantities   are calculated if the states $\alpha,\beta$
 are defined 
 using  wave packets \cite{Ishikawa-Shimomura}. The wave packet
is  defined around a certain time and position and has the energy and
momentum of finite uncertainties and is useful
for studying the space time dependent informations.     
We study the amplitudes of the wave packets and find  the finite
  time or length dependences of the amplitudes. 

First  the
  coherence lengths in which the waves keep coherences   
  are evaluated from the production and detection processes. 
Neutrinos are produced by a decay of particles which have certain  
coherence lengths.   As was analyzed in Ref.\cite{Ishikawa-Tobita-ptp}, the coherence lengths
are the sizes of waves that maintain coherence. The mean free path, on
the other hand, is calculated from the transition probability and the
density of scatterers as the average distance that a particular wave
propagates freely.  Hence  the coherence lengths are determined 
by mean free paths. 
 
(1-1)Decay of pion in fright.

Pion is produced in matter, in metal for instance,  and the mean free
path was 
calculated  \cite{Ishikawa-Tobita-ptp} as
\begin{eqnarray}
\sigma_{pion}=10^{-1}[\text{m}].
\end{eqnarray}
When these pions are emitted from metal into the vacuum, these waves 
maintain coherence within this size. So the length in which particles
maintain coherence is determined by the mean free path. These 
particles are described by the   wave packets of the finite coherence
length in the vacuum. 

(1-2)Decay of Nucleus  in solid. 

The other sources of neutrino are  unstable nucleus. In reactor, these unstable
nucleus are bound in atoms, and they are described by wave functions of
finite sizes. The sizes of nucleus wave functions are estimated from
the center of mass gravity effect between the nucleus and electrons in
the atom \cite{Ishikawa-Tobita-ptp} as, 
\begin{eqnarray}
\sigma_{Nucleus}=r\times R_{atom}={1 \over 2000}\times 10^{-10} \approx 10^{-13}[\text{m}],
\label{nucleus-size}
\end{eqnarray}
where $r$ is the ratio between the mass of nucleus and electrons and
$R_{atom}$ is the size of atom. The size of nucleus wave packet is in
the range between the nucleus size and atom size. So, the
size of pion wave packet in fright is large but the size of nucleus wave 
packet is small. 


Neutrinos interact with particles in the target such as electrons or
nucleus which have finite coherence lengths. The electron has a large
size but the nucleus has a small size, and they are defined by wave
packets.

(2-1)Electron wave function in matter.

Bound electrons in atom are described by atomic wave functions.
The wave packet of the electron  has a size of the electron wave
function 
\begin{eqnarray}
\sigma_{electron}=10^{-10}[\text{m}].
\end{eqnarray}

(2-2) Nuclear wave function  in solid  

Nucleus is extremely small and is described by a nucleus wave
function. From the size of nucleus wave function
Eq.$(\ref{nucleus-size})$, the 
wave packet of nucleus    
has a size 
\begin{eqnarray}
\sigma_{Nucleus}=10^{-13}[\text{m}],
\label{nucleus-size2}
\end{eqnarray}
which is the same as Eq.~($\ref{nucleus-size}$).  Thus the nucleus has a
small size.

Wave packet is a superposition of plane waves with a momentum dependent 
weight. Gaussian and Lorentzian wave packets are  studied often. They
become maxima around the central region and decrease
uniformly at large momentum region. The former  
decreases rapidly but the latter decreases slowly. In the central region
, Gaussian wave packet
is most convenient for  practical calculations and is applied here also.
In  the  tail region, where the momentum becomes large, the
particle correlations  are determined here  from their 
production processes.

The tail of the wave packet of pion that is produced in hadron
collisions depends on the production processes. They are described by
pion correlation functions 
\begin{eqnarray}
\Delta_{p}(x_1-x_2)=\langle \alpha|\varphi(x_1)|\beta \rangle \langle \beta |\varphi(x_2)|\alpha \rangle,
\end{eqnarray}  
for the initial and final state $|\alpha \rangle,| \beta \rangle$ and is 
transformed as     
\begin{eqnarray}
\Delta_{p}(x_1-x_2)=\int d^4q e^{ip(x_1-x_2)} \Delta_p(q).
\end{eqnarray}
The $\Delta(q)$ of a free relativistic particle, $\Delta_p^{(free)}(q)$,
 and of an interacting particle with $\varphi^4(x)$
 interaction, $\Delta_p^{(int)}(q)$, are  
\begin{eqnarray}
& &\Delta^{(free)}(q)={1 \over (2\pi)^4}{1 \over q^2-m^2-i\epsilon},\\
& &\Delta^{(int)}(q)={1 \over (2\pi)^4}{1 \over q^2-m^2-i\epsilon}{1 \over
 q^2-m^2+i\epsilon}\delta((q+q_{\delta})^2-m^2).
\end{eqnarray}
From these correlation functions, $\Delta_p^{(free)}(x)$ and $\Delta_p^{(int)}(x)$ are calculated in
Euclidean metric in the limit $\lambda=\sqrt{x^2} \rightarrow 0$ as
\begin{eqnarray}
& &\Delta_p^{(free)}(x)={1 \over 4\pi}\delta(\lambda) + less~ singular~ term
\label{free-correlation},\\
& &\Delta_p^{(int)}(x)={1 \over 64 \pi^2}+O(\lambda),~:~q_{\delta} \rightarrow 0.
\label{interactig-correlation}
\end{eqnarray}

\section{Wave packet evolution}

A wave packet is a superposition of the momentum states and has a finite
spatial width \cite{Goldberger}\cite{newton}\cite{Sasakawa}\cite{ishikawa}. Although a  neutrino wave packet is an intermediate state in
our formalism, it is
instructive and useful to analyze the wave packet and its evolution in
time here. The wave packet   behaves like a particle and has the center 
position of momentum, space, and time. Although it behaves like a
particle,  it also behaves like a wave and has  a characteristic
phase. So the wave packet is not the same as 
an ordinary particle. 
 
Wave packet of the central values of the position, ${\vec X}_0$, the  
momentum, ${\vec P}_0$, and the time, ${\vec T}_0$   
is  decomposed as 
\begin{eqnarray}
& &\langle t,{\vec x}|{\vec P}_0,{\vec X}_0,T_0 \rangle = \int d{\vec p} \langle {\vec x}|{\vec p}\rangle \langle t,{\vec p}|{\vec P}_0,{\vec X}_0,T_0 \rangle\\
 & &= N_3 \left({\sigma \over 2\pi}\right)^{\frac{3}{2}} \int d{\vec p} e^{-iE({\vec p})(t-T_0)+ i{\vec p}\cdot( {\vec x}-{\vec X}_0)-{\sigma \over 2}({\vec p}-{\vec P}_0)^2}, \nonumber
\end{eqnarray}
where $\sigma$ is the size of wave packet. The wave packet maintains
its shape and size   during  small $t-T_0$ period, and it has a
translational motion.
\begin{eqnarray}
\label{eq:closetime}
& &\langle t,{\vec x}|{\vec P}_0,{\vec X}_0,T_0 \rangle =N_3 e^{-{1 \over 2 \sigma}({\vec x}-{\vec X}_0-{\vec v}_0(t-T_0))^2}
e^{-i \phi},
\\
& &\phi=E({\vec P}_0)(t-T_0)-{\vec P}_0 \cdot ({\vec x}-{\vec
 X}_0),~ {\vec v}_0={p_i \over E({\vec p})}.
\end{eqnarray}  
 
The wave is extended around the center ${\vec x}(t)={\vec X}_0+{\vec
v}_0(t-T_0)$, and is classified by the phase $\phi$. The phase 
at the center is given by
\begin{eqnarray}
{ \phi=E({\vec P}_0)(t-T_0)-{\vec P}_0 \cdot ({\vec x}-{\vec
 X}_0)=\left(E-{p^2 \over E}\right)(t-T_0)={m^2 \over E}(t-T_0)}.
\end{eqnarray}
This phase is proportional to the mass squared and inversely proportional
to the energy. Hence its magnitude becomes extremely small. 
At   large $t-T_0$, the stationary momentum that satisfies  
\begin{equation}
{\partial \over \partial p_i} \left({-iE({\vec p})(t-T_0)+ i{\vec p}\cdot( {\vec x}-{\vec X}_0)-{\sigma \over 2}({\vec p}-{\vec P}_0)^2} \right)=0.
\end{equation}
gives a dominant contribution to the above momentum integral. 
The momentum which satisfies this equation is determined by the time and
space coordinates, 
\begin{eqnarray}
& &{\vec P}_X={\vec P^{(0)}}_X+O\left({1 \over t-T_0}\right), \\
& &{\vec P^{(0)}}_X= m{1 \over \sqrt{(t-T_0)^2-({\vec x}-{\vec X}_0)^2 } }({\vec x}-{\vec X}_0)
\end{eqnarray}
and the wave function is described as 
\begin{align}
\label{eq:asymptotic}
 &\langle t,{\vec x}|{\vec P}_0,{\vec X}_0,T_0 \rangle \\
 &= N_3 e^{-iE({\vec p})(t-T_0)+ i{\vec p}\cdot( {\vec x}-{\vec X}_0)  }\left({1 \over 2i{ \gamma_L \over \sigma}+1}\right)^{\frac{1}{2}}\left({1 \over 2i{ \gamma_T \over \sigma}+1}\right)  e^{-{1 \over 2 }\sigma ({{\vec P^{(0)}}_X}-{{\vec P}_0})^2 },  
\end{align}
This wave has    the same phase 
\begin{eqnarray}
 \phi={m^2 \over E}(t-T_0)
\end{eqnarray} 
as
before and the magnitude varies with time. The magnitude decreases and
the wave expands with time in space. The expansion  parameters in the
longitudinal direction, $\gamma_L$, and in the transverse direction, $\gamma_T$  are 
\begin{eqnarray}
\gamma_L={1 \over 2}{m^2 |t-T_0| \over E^3({\vec P}_X)}, ~\gamma_T={1 \over 2}{|t-T_| \over E({\vec P}_X)}.
\end{eqnarray}
The longitudinal expansion is proportional to the mass squared and
vanishes in the massless case. The transverse expansion is independent
from the mass and is determined by the initial size of the wave packet.

\section{Probability at finite time interval}

{\bf Probability at finite time interval :T2K near detector}

The transition amplitude of the neutrino reaction which includes the
production and detection processes are studied by wave packets 
\cite{neutrino-wave-packet}. The
neutrino is treated as a wave in the intermediate states  and its size
is determined by the particle and wave properties  in the initial and 
final states.

The amplitude $T$ for  a T2K neutrino reaction that includes 
the production and propagation processes where  the neutrino  is in the 
outgoing state is given as 
\begin{eqnarray}
T &=& ig m_{\mu} N' \int d t d{\vec x} d{\vec p}_{\nu}
\langle\beta_f|\varphi(x)|\alpha_i \rangle \left(\frac{m_{\nu}}{E(\vec{p}_{\nu})}\right)^{\frac{1}{2}}\label{amplitude}\\
& & \times e^{i\left(E(\vec{p}_{\mu})t-{\vec p}_{\mu}\cdot{\vec x}\right)}\bar{u}({\vec p}_{\mu}) \gamma_5 \nu({\vec p}_{\nu})e^{i\left(E({\vec p}_{\nu})(t-T_{\nu})-{\vec p}_{\nu}\cdot({\vec x}-{\vec X}_{\nu})\right) -{\sigma_{\nu} \over 2}({\vec p}_{\nu}-{\vec k}_{\nu})^2},\nonumber\\
N' &=& \frac{N_{\nu}}{(2\pi)^{\frac{3}{2}}} \left(\frac{m_{\mu}}{E(\vec{p}_{\mu})}\right)^{\frac{1}{2}},\nonumber\\
\langle \beta_f |\varphi(x)|\alpha_i \rangle &=& \left(\frac{4\pi}{\sigma_{\pi}}\right)^{\frac{3}{4}}e^{-i\left(E({\vec
 p}_{pion})(t-T_{\pi})-{\vec p}_{pion}\cdot({\vec x}-{\vec
 X}_{\pi})\right)}\nonumber\\
& & \times e^{-{1 \over 2 \sigma_{\pi} }\left({\vec x}-{\vec X_{\pi}}-{\vec
 v}_{\pi}(t-T_{\pi})\right)^2}T_{\beta'_f,\alpha_i}.
\end{eqnarray}
In the above equation, $|\alpha_i \rangle $ is the initial state and
$|\beta_f \rangle $ is the final state.  $\varphi(x)$ is the pion field and
$(t,{\vec x})$ is  the space time coordinates where  the weak
interaction takes place. The pion is expressed by the wave packet of the
size $\sigma_{\pi}$  and is produced at the space time coordinate
${(T_{\pi},{\vec X}_{\pi})}$. The neutrino is detected with the wave
packet of the size $\sigma_{\nu}$ and is detected at  
$(T_{\nu},{\vec X}_{\nu})$.

After the tedious calculations, we have integrated probability of
observing neutrino at a 
finite time T as,
\begin{eqnarray}
& &\int d{\vec p}_\text{muon} d{\vec x}_{neutrino} d{\vec p}_{pion}\sum_{s_1,s_2}|T|^2 \\
& &= N \int_0^{\text{T}} dt_1 dt_2  {e^{i {m_{\nu}^2 \over E_{\nu}}   (t_1-t_2) }
 \over (t_1-t_2)} \left(1 +
		   f_\text{normal}e^{i\omega_{\pi}(t_1-t_2)}\right)
\label{probability2} \nonumber \\
& &=
\frac{\tilde{N}_{\text{prob}}}{E_{\nu}} ~\left[
					  F_\text{universal}(0)\times g(\text{T},\omega_{\nu}) + f_{\text{normal}}\times g(\text{T},\omega_{\pi}
	   + \omega_{\nu})\right],\nonumber
\end{eqnarray}
\begin{eqnarray}
\tilde N_{\text{prob}} &=& 
(2\sqrt{2}\pi)^3
\sigma_{\nu}
 f_{\mu},~
f_{\mu} = g^2m_{\mu}^2(m_{\pi}^2 - m_{\mu}^2),\nonumber \\
g(\text{T},\omega) &=& \text{T} \int^\text{T}_0 dX \frac{\sin \left(\omega X\right)}{X} +
  \frac{1}{\omega}\left\{\cos(\omega \text{T}) - 1\right\},\label{oscilating-term}\\
\omega_{\nu}&=&\frac{m_{\nu}^2}{E_{\nu}},~\omega_{\pi}=\frac{m_{\pi}^2}{E_{\pi}},~L=c\text{T},\nonumber
\end{eqnarray}
\begin{figure}[t]
 \centering
 \includegraphics[angle=-90, scale=.4]{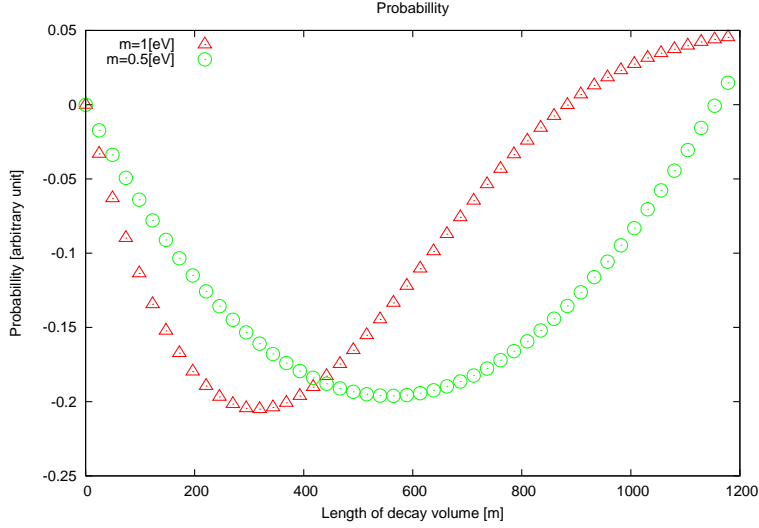}
 \caption{The length dependence of the oscillating part in the 1 [GeV] neutrino 
detection probability of the mass  $1\, [\text{eV}/c^2]$ and $0.5\, [\text{eV}/c^2]$ is shown.}
\label{Finite-length-probability }
\end{figure}
where the coefficients $F_\text{universal}$ is determined from the large
momentum contribution at the tail of the wave packet
Eqs.$(\ref{interactig-correlation})$ and $f_\text{normal}$ is
determined from the states at the central region of the wave packet.
The probability at a finite time T has a T-linear term and an
oscillating term.


The oscillating term for the mass $1\,[\text{eV}/c^2]$ and $0.5\,[\text{eV}/c^2]$ is
given in Fig. (\ref{Finite-length-probability }). The probability
becomes minimum at around $300\,[\text{m}]$ for $1\,[\text{eV}/c^2]$ and at around $600\,[\text{m}]$ for
$0.5\,[\text{eV}/c^2]$ and is expected to be observable.

{\bf  KamLAND neutrino:flavour oscillation}

The neutrino production and detection   processes are the second order weak
process at the coordinate $x_1^{\mu}$ and the 
detection of the coordinate $x_2^{\mu}$. The particle states $\alpha_i$
and $\beta_i$ in the
initial state are defined by the wave packets and integration on the 
variables,  
${\vec x}_1$ and $x_2^{\mu}$ are made easily.  The amplitude for 
KamLAND neutrino becomes, then, 
\begin{eqnarray}
T &=& iN'  \int d x_1^0 T_1 \bar{u}({\vec p}_{l_i})(1-\gamma_5)
\gamma^{\mu_1}\int d{\vec p}_{\nu} {\gamma p+m_{\nu}  \over 2 E({\vec p}_{\nu})}
e^{-ip_{\nu}(X_1-X_2)} 
\label{amplitude4} \\
&\times& e^{-\sigma_{f}^p({\vec p}-{\vec P_{f}})^2}
 e^{i( {p^0}-{ P_{\alpha}^0} )x_1^0} 
 e^{-\sigma_f^0({p^0}-{ P_{\beta}^0})^2} \Gamma u({\vec p}_{l_f}) \tilde T_2, 
\label{Kam-land-formula}  \nonumber
\end{eqnarray} 
where $X_1^{\mu}$ and $X_2^{\mu}$ are the space and time coordinates of 
the particles at production and detection and $T_1$ and $T_2$ are the
amplitudes where the neutrino is produced or detected.  The integrand is an
exponential function of those that have real parts and imaginary parts and
the integral is obtained either  around the minimum of the real
part, Gaussian point, ${\vec k}_{\nu}$, or the minimum of the imaginary part, stationary
phase point.  The wave packet sizes $\sigma_f^p$  and $\sigma_f^0$ are
defined from the wave packet sizes of particles. The  Gaussian 
approximation  is good at small time
and the stationary phase approximation   is good at large time.


{\bf  Gaussian point}

Using integral around Gaussian point, the neutrino in the above amplitude is described
by a wave function 
$T_0({\vec k}_{\nu},X_1-X_2)$ that is  described as  
\begin{eqnarray}
& &T_0({\vec k}_{\nu},X_1-X_2)= \exp{(i\phi(k_{\nu}))} \tilde T({\vec k}_{\nu},X_1-X_2),            \\
& &\tilde T({\vec k}_{\nu},X_1-X_2)=\exp\left[-\tilde \sigma_x^T(({\vec X}_1-{\vec
 X}_2)^T)^2-\tilde \sigma_x^L(({\vec X}_1-{\vec
 X}_2)^L-{\vec v}(X_1^0-X_2^0))^2\right], \nonumber
\label{oscillation-amp}\\
& &\tilde \sigma_x^T={1 \over 2\sigma_T } ,\tilde \sigma_x^L= {1 \over 2\sigma_L }, \nonumber
\end{eqnarray}
where $\phi(k_{\nu})$ is the phase of the intermediate neutrino.

{ \bf (P-detection\,:\,large $\sigma_L$ )} 

When $\sigma_L$ is large, the variable $X_1^0-X_2^0$ is extended 
in wide area and is
integrated with the phase $e^{iE({\vec k}_{\nu}-E_{\nu})(X_1^0-X_2^0)}$. 
Then,  we have $E({\vec k}_{\nu})=E_{\nu}$ and the phase of the amplitude 
becomes 
\begin{eqnarray}
\phi&=&{\vec k}_{\nu}\cdot({\vec X}_1-{\vec
 X}_2)\label{phase2}\nonumber \\
&=& E_{\nu} |({\vec X}_1-{\vec X}_2)|+\phi(m) \nonumber ,\\
\phi(m)&=&{{m_{\nu}}^2 \over 2 E_{\nu}}
 |({\vec X}_1-{\vec X}_2|,
\label{phase-P-detection}
\end{eqnarray}
which agrees to the standard formula of the phase of a momentum state.
The wide wave packet is approximated well with the plane wave and the
oscillation phase of the wave packet agrees to that of the plane wave\cite{Lipkin}.    

{ \bf  ( X-detection\,:\,Small $\sigma_L$ )}

  In the small $\sigma_L$, the wave function becomes finite only
  in a narrow region around the position, 
 \begin{eqnarray}
{\vec X}_1={\vec X}_2+{\vec v}_{\nu}(X_1^0-X_2^0),
\end{eqnarray}  
and the  time $x^0$ integration   in Eq.$(\ref{Kam-land-formula})$ is
made by substituting this relation to Eq.$(\ref{Kam-land-formula})$ 
 and the phase $\phi$  is rewritten as    
\begin{eqnarray}
\phi(m)&=&-E({\vec k}_{\nu})(X_1^0-X_2^0)+{\vec k}_{\nu}\cdot({\vec
 X}_1-{\vec X}_2)\label{phase}\nonumber \\
&=&(-E({\vec k}_{\nu}){1 \over v_{\nu}}+k_{\nu})|{\vec X}_1-{\vec X}_2|=
 -{{m_{\nu}}^2 \over k_{\nu}}|{\vec X}_1-{\vec X}_2| ,
\label{phase-X-detection}
\end{eqnarray}

Thus the phase of amplitude for the small wave packet is given in Eq.$(\ref{phase-X-detection})$ 
and that for the large wave packet is   given in $(\ref{phase-P-detection})$  . They have different forms. 
In flavour oscillation, the constant phase which does not depend on the
neutrino mass cancels and the mass dependent phase $\phi(m)$ determines
the interference term. The  mass dependent phase $\phi(m)$ is described
by a unified formula, 
\begin{eqnarray}
 \phi(m)= - {{m_{\nu}}^2 \over \gamma  E_{\nu}(\vec{k}_{\nu})} |{\vec
  X}_1-{\vec X}_2|,  
\end{eqnarray}
with a new parameter $\gamma$. $\gamma=2$ is the value of wide wave
packet and agrees to that of the standard oscillation formula 
discussed usually and $\gamma=1$ is the value of the narrow wave packet
and reveals a phase of the relativistic particle \cite{Kayser}.   

{\bf Stationary phase approximation}

In Eq.$(\ref{amplitude4})$, the momentum integration is made using the
stationary phase approximation. The stationary momentum,
$p_{\nu}^{(0)}$, for $\sigma=0$ is 
\begin{eqnarray}
{\vec p}_{\nu}^{\,(0)}=m{({\vec x}-{\vec X}_{\nu})
\over((x^0-X_{\nu}^0)^2-({\vec x}-{\vec X}_{\nu})^2)^{\frac{1}{2}} }
\end{eqnarray} 
is proportional to the mass. After the time $x^0$ integration is made, 
the amplitude becomes 
\begin{eqnarray}
T_0({\vec k},X_1-X_2)&=&  e^{-i {m_{\nu}^2 \over E_{\nu}} (x^0-X_{\nu}^0)}e^{-
{\sigma_{\nu} \over 2}\left({\vec
p}_{\nu}^{~(0)}-{\vec k}_{\nu}\right)^2}, \label{amplitude2}
\end{eqnarray}
which is the same as the above X-detection Eq.~$(\ref{phase-X-detection})$.

{\bf Transition from P-detection to X-detection}

Since the oscillation formula for the narrow wave packet is different
from the wide wave packet, it is worthwhile to know the wave packet size 
of the boundary.  In a detection where the wave packet size $L_{w.p}$ is
equivalent to  de Broglie  wave length $\lambda_{w}$, the X-detection should be better than the
P-detection. So naively  the boundary between two detections is
expected at 
\begin{eqnarray}
L_{w. p}=\lambda_{w}
\end{eqnarray}
The wave packet sizes for the KamLAND are  given in
Eq.$(\ref{nucleus-size})$ and 
$(\ref{nucleus-size2})$. Using these values we find  
from a numerical study of the integration of Eq.$(\ref{amplitude})$  that
    the transition from the $\gamma=2$ to $\gamma=1$ occurs at
$E_{\nu}=2\,[\text{MeV}/c]$. This value is the end region of KamLAND neutrino
detections and the previous analysis of KamLAND will not be affected. In
other neutrino detectors that use electrons, the wave
packets are much larger than the de Brogle wave length for the neutrino 
energy in a few $[\text{MeV}/c^2]$, hence the detection are regarded as
P-detection. Its oscillation length is that of the standard formula.
 
At the interface  region between the X-detection and P-detection where
the amplitude
becomes a superposition of those  of $\gamma=1$ and $\gamma=2$,
\begin{eqnarray}
& &f(m_i)=f_1 (m_i)+f_2 (m_i)\\   
& &f_1(m_i)=e^{i{m_i^2 \over E}t},f_2(m_i)=e^{i{m_i^2 \over 2E}t},\nonumber
\end{eqnarray}    
the probability of flavour oscillation of the masses $m_1$ and
$m_2$ in this region behaves  as 
\begin{eqnarray}
& &P=\left|\sum_i ( f_1(m_i)+f_2(m_2))\right|^2\\
& &=\sum_{ij}\left( e^{i{m_i^2-m_j^2 \over E}t}+e^{i{m_i^2-m_j^2 \over 2
 E}t}+e^{i{2m_i^2-m_j^2 \over 2 E}t}+e^{i{m_i^2-2m_j^2 \over 2
 E}t}\right)\nonumber \\
& &= 2+2\cos\left({\delta m^2 \over E}t\right)+ 2\cos\left({\delta m^2 \over 2
 E}t\right)+2\cos\left({\delta m^2 +m_1^2\over E}t\right)+ 2\cos\left({\delta m^2 +m_1^2 \over
 2 E}t\right),
 \nonumber
\end{eqnarray}  
where $\delta m^2=m_1^2-m_2^2$.
The oscillation is determined by the mass-squared difference
$\delta m^2 $ and the absolute value of one  
mass $m_1^2$. For $\delta m^2 \ll m_1^2$, the oscillation period is
determined by the absolute value of the mass $m_1$ in this region, and
 for $\delta m^2 \approx m_1^2$ the oscillation becomes complicated functions
 of the time and energy.  
\section{Results and summary}
Transition amplitudes of neutrinos where the production and
detection processes are fully taken into account are studied. This 
amplitudes show that the
macroscopic interference of the neutrino wave is possible. The
probability at the
finite distance show an unusual oscillation that is caused by the 
interferences of neutrino wave produced at different positions.   Using
T2K near detector,  this new oscillation might be observable if the mass
of the heaviest neutrino is around $0.5-1~[\text{eV}/c^2]$. 

This  amplitude is applied to KamLAND neutrino and 
generalized oscillation formula,
\begin{eqnarray}
T_0({\vec k}_{\nu},X_1-X_2)&=&  e^{-i{ {m_{\nu}^2 \over \gamma E_{\nu}}} (x^0-X_{\nu}^0)}e^{-i{
{\sigma_{\nu} \over 2}} \left({\vec
p}_{\nu}^{~(0)}-{\vec k}_{\nu}\right)^2}, \label{amplitude2}
\end{eqnarray}
is obtained for the neutrino amplitude. The  $\gamma$ depends
on several parameters and a transition from $\gamma=1$ to $\gamma=2$ is 
expected at $E_{\nu}=2\,[\text{MeV}]$. 

\section*{Acknowledgements}
One of the authors (K.I) thanks Dr.~Nishikawa for useful discussions on 
the near detector of T2K experiment, Dr.~Inoue on  KamLAND detector,
Dr.~Asai, Dr.~Mori, and Dr.~Yamada for useful discussions on interferences. This work was partially supported
by a Grant-in-Aid for Scientific Research(Grant No. 19540253 ) provided
by the Ministry of Education,
Science, Sports and Culture,and a Grant-in-Aid for Scientific Research on 
Priority Area ( Progress in Elementary Particle Physics of the 21st
Century through Discoveries  of Higgs Boson and Supersymmetry, Grant 
No. 16081201) provided by 
the Ministry of Education, Science, Sports and Culture, Japan. 
\\
{}
\appendix
\def\thesection{Appendix \Alph{section}}
\def\thesubsection{\Alph{section}-\Roman{subsection}}

\end{document}